\documentclass[pra,superscriptaddress,showpacs]{revtex4}
\usepackage{graphicx}       
\usepackage{bm}         
\usepackage{amsmath}        
\usepackage{latexsym}
\begin{document}


\title{Comparing the states of many quantum systems}
\author{IGOR JEX$^1$, ERIKA ANDERSSON$^2$, and ANTHONY CHEFLES$^3$\\
$^1$Department of Physics, FNSPE, Czech Technical University Prague,\\
B\v rehov\'a 7, 115 19 Praha, Czech Republic,\\
e-mail igor.jex@fjfi.cvut.cz\\
$^2$Department of Physics, University of Strathclyde, Glasgow G4 0NG, UK,\\
tel: +44 141 5483376, fax: +44 141 5522891,\\
 e-mail erika@phys.strath.ac.uk \\
$^3$Department of Physical Sciences, University of Hertfordshire,\\
Hatfield AL10 9AB, Hertfordshire, UK, \\
e-mail anthony.chefles@herts.ac.uk}
\date{\today}
\begin{abstract}
We investigate how to determine whether the states of a set of quantum systems are identical or not. 
This paper treats both error-free comparison, and comparison where errors in the result are allowed. Error-free comparison means that we aim to obtain definite answers, which are known to be correct, as often as possible. In general, we will have to accept also inconclusive results, giving no information. To obtain a definite answer that the states of the systems are not identical is always possible, whereas, in the situation considered here, a definite answer that they are identical will not be possible. 
The optimal universal error-free comparison strategy is a projection onto the totally symmetric and the different non-symmetric subspaces, invariant under permutations and unitary transformations. 
We also show how to construct optimal comparison strategies when allowing for some errors in the result, minimising either the error probability, or the average cost of making an error.
We point out that it is possible to realise universal error-free comparison strategies using only linear elements and particle detectors, albeit with less than ideal efficiency. Also minimum-error and minimum-cost strategies may sometimes be realised in this way. This is of great significance for practical applications of quantum comparison. 
\end{abstract}
\pacs{03.67.-a, 03.65.Ta, 42.50.Dv}
\maketitle

\section{Introduction}

When comparing classical systems, a straightforward way to proceed is to measure observables of each system individually, and then compare the measurement results. 
To compare the states of quantum systems is less straightforward, since results of quantum measurements are statistical in their nature. In addition, the measurement usually introduces a nonnegligible disturbance of the measured state. In quantum mechanics, simultaneous measurements of non-commuting observables are restricted. If only a single copy of each quantum system is available, we cannot measure all observables of the systems precisely, and thus cannot compare the states of quantum systems in the same way as classical systems. Based on measurements of the individual quantum systems, it is in general only possible to make statistical predictions of their similarities and differences. If an ensemble of identically prepared systems is available for each of the compared quantum states, the result will be more reliable.

Quantum mechanics, however, allows us to perform collective measurements, which sometimes have no classical analogue. An example is the projection onto entangled Bell states, used in classically impossible tasks such as dense coding \cite{ben}. Using collective measurements, it is possible to reliably compare many quantum systems, even when only a single copy of each system is available. 
In this paper we will investigate how to determine whether the states of a set of quantum states are identical or different.  We first consider the situation where the obtained results are {\it unambiguous}, meaning that, when an answer is obtained, for example, that the states of the systems are different, it is always true. This forces us to accept that the comparison may sometimes give an inconclusive answer, which means that the attempt to determine whether the states are identical or not has failed. The error-free character of the answers is useful whenever one wants to be absolutely certain that it is correct. 

Dropping the requirement that the result of the comparison has to be error-free, we also show how to construct optimal comparison strategies giving the least possible error (or least possible cost of error) in the result. With such a strategy, no matter how many times the comparison is repeated on new copies of the quantum systems, one can only estimate that the probability that the answer is wrong is below a certain limit. To reach the limit, many repetitions may be required, whereas an error-free strategy sometimes gives a definite answer in only one try. On the other hand, the error-free strategy will often have a large probability to fail, and from this point of view we may prefer a  minimum-error or minimum-cost comparison strategy.

Quantum comparison of two pure quantum systems has recently been considered in \cite{steve}. It is also possible to compare unitary transforms to each other \cite{ucomp}, a task which is related to state comparison. Quantum comparison may be used in many quantum information applications, such as error correction, 
or for checking that copies of `keys' are identical and go into their `locks'. Small effects may be monitored by comparison with a highly reliable reference system. Two recent  applications where quantum comparison is needed are quantum fingerprinting \cite{buhr} and quantum digital signatures \cite{chuang}. Because of this, it is important to understand the full possibilities offered, and the absolute limitations of quantum comparison, and to investigate how quantum comparison could be experimentally realised. 

The universal error-free comparison strategy is investigated in section \ref{secuni}.
It is optimal when the distribution of possible quantum states is flat, and allows us to unambiguously (without error)  detect a difference in the compared quantum systems. It is not possible to unambiguously determine that all the states are identical, unless we have more information about the states we want to compare \cite{tonyprivate}.
It is possible, however, to obtain more information about differences among the $N$ quantum systems, for example, that no $M$ particles, where $M\leq N$, have identical states.  This `full universal error-free comparison strategy' will be introduced in section \ref{subspaces}. In section \ref{prob}, we show how to obtain another type of optimal comparison strategies, minimising either the error in the result, or the average cost of making an error. The optimal strategies can be constructed for comparing both pure and mixed quantum states, and noise and entanglement with an environment can be taken into account.  
We also investigate, in section \ref{realisations}, how comparison strategies may be realised. 
We find that  universal error-free comparison can be effected using only beam splitters and particle detectors, making it possible to realise using linear optics. This is of great importance for experimental realisations. Also minimum-error and minimum-cost comparison can in principle be realised as a projective measurement in some basis, but the realisation depends very much on the individual situation.
We end with a discussion and conclusions.

\section{Universal error-free comparison strategy}
\label{secuni}

Let us suppose that we have a collection of quantum systems.
Is it possible to compare the states of these quantum systems, that is, to determine whether the states are all identical or not, without necessarily obtaining any information about the individual states?
In this section we will investigate error-free quantum comparison, meaning that whenever one obtains the result `same' or `different', it is known also to be correct. This will force us to accept that  we sometimes get an inconclusive result, meaning no result at all. For reasons we will return to later, all the states are, in this section, assumed to be pure. We want to find the optimal comparison procedure when the quantum states are completely unknown, in other words, when the distribution of possible states for all the particles is flat. We refer to this as `universal comparison'.   It is important to note that the resulting optimal universal comparison strategy will give correct answers also when there {\it is} prior information about the states, even if it in that case may not be optimal, in the sense that it may not give unambiguous answers as often as possible.

The optimal way to compare, without error, the states of two unknown, 
pure quantum systems, is to check whether their combined state is symmetric or antisymmetric \cite{steve}. 
If the two quantum systems have been prepared in the same state, then their combined state is necessarily symmetric. From this follows that finding the systems in an antisymmetric state definitely indicates that the two systems were different. On the other hand, finding the two systems in a symmetric state does not necessarily mean that they were identically prepared, but rather gives us no definite information. (We are comparing the internal states, assuming that the quantum systems may be distinguished at least through different spatial positions; their total state is of course symmetric (bosons) or antisymmetric (fermions).) 

Also when comparing the states of more than two quantum systems, if they are all identically prepared, then their overall state will be symmetric. This means that finding the $N$-particle system in the non-symmetric part of its total Hilbert space indicates that the subsystems cannot all have been identically prepared. 
The symmetric/non-symmetric projection is the optimal comparison strategy when we are asking for an error-free answer. When there is no prior information about the states of the particles, the best comparison strategy must be invariant under permutations of, and unitary transforms on the compared quantum systems (the same transform on each system). Physically, the permutation invariance comes from the fact that we are not asking which of the individual quantum states are different from each other, and the  invariance under unitary transformations arises because we do not know in which basis they are given (in fact we do not know the states at all).
This leads us to consider the corresponding invariant subspaces, which
are the symmetric and non-symmetric subspaces. 
In the non-symmetric subspace, the quantum systems are never all in the same internal state. In 
the symmetric subspace, the states may be identical or different, corresponding to an 
inconclusive outcome. There is no invariant subspace in which the quantum states always are
identical, and therefore we can never obtain an error-free answer that the states were all identical.

To further motivate the optimality, we may think of the comparison as a generalised measurement with three outcomes, `yes, all the states are the same', `no, all the states are not the same', and `no certain answer'. To these outcomes correspond three measurement operators $\Pi_Y, \Pi_N$ and $\Pi_?$, respectively. The probability of obtaining the different results are given by
\begin{equation}
p(Y)=\langle\psi|\Pi_Y|\psi\rangle,\quad
p(N)=\langle\psi|\Pi_N|\psi\rangle,\quad
p(?)=\langle\psi|\Pi_?|\psi\rangle,
\end{equation}
where $|\psi\rangle$ is the many-particle state of the system. The measurement operators $\Pi_Y, \Pi_N$ and $\Pi_?$ have to be positive, meaning that they have only nonnegative eigenvalues. This guarantees that all the probabilities $p(Y), p(N)$ and $p(?)$ are nonnegative for any $|\psi\rangle$. In addition, $\Pi_Y+\Pi_N+\Pi_?=\mathbf{1}$ has to hold, corresponding to $p(Y)+p(N)+p(?)=1$.  The measurement operators do not have to be pure state projectors. Generalised measurements are usually referred to as POM (probability operator measure) or POVM (positive operator-valued measure) strategies \cite{hel}.

For error-free comparison, we demand that $p(Y)=0$ whenever the states of the systems were not identical, and $p(N)=0$ whenever they were all identical. This means that the expectation value of $\Pi_N$ for any state in the totally symmetric subspace must be zero. In the optimal comparison measurement, to make $p(?)$ as small as possible, the support of the operators $\Pi_Y$ and $\Pi_N$ should include as much of the total Hilbert space as possible. The optimal $\Pi_N$ is therefore seen to be the projector onto all of the non-symmetric subspace. 
The invariance under permutations and unitary transformations applying to the comparison situation, is thus seen to correspond to the measurement operators possessing the same invariances. 
Since there is no invariant subspace where the states are always identical, $\Pi_Y=0$, and $\Pi_?$ is the projector onto the symmetric subspace.
To summarise, for the optimal comparison strategy we have
\begin{equation}
\Pi_N=P_{nonsym} , \quad\Pi_?=P_{sym},  \quad\Pi_Y=0.
\end{equation}
Let us here also clarify that asking questions about the phases of the individual quantum states or systems is not meaningful, since the total combined quantum system only has an overall phase of its wave function.

The average probability of detecting the states of the systems as different may be understood as the average success rate of the comparison. This probability is given by the average expectation value of $P_{nonsym}$, where the average is taken over all possible states $|\Psi_N\rangle$ of the $N$-particle system. Since there is no prior information about the individual states, their distribution is flat and the  integral $\int|\Psi_N\rangle\langle\Psi_N| d\Psi_N$ is equal to the identity operator in the total $N$-particle Hilbert space. The average success probability is therefore 
\begin{eqnarray}
\label{successintegral}
p_{success}&=&\overline{\langle P_{nonsym}\rangle}=
\int\langle\Psi_N|P_{nonsym}|\Psi_N\rangle  d\Psi_N\nonumber\\
&=&Tr\{P_{nonsym}\int|\Psi_N\rangle\langle\Psi_N| d\Psi_N\}
=Tr\{ P_{nonsym} \otimes \mathbf{1}^N\}\nonumber\\
&=&{D_{nonsym}\over D_{tot}}=1-\overline{\langle P_{sym}\rangle}=1-{D_{sym}\over D_{tot}},
\end{eqnarray}
where $D_{sym}$ and  $D_{nonsym}$  are the dimensions of the symmetric and nonsymmetric subspaces, and $D_{tot}=D^N$ is the dimension of the total Hilbert space (the particles may be distinguished since they are spatially separated; their total quantum state including the spatial part will be symmetric or antisymmetric depending on whether they are bosons or fermions). The dimension of the symmetric subspace for $N$ quantum systems in $D$ dimensions is equal to the number of ways we can distribute $N$ particles in $D$ different boxes, which is
\begin{equation}
D_{sym}={D+N-1\choose D-1}={D+N-1\choose N}.
\end{equation}
For each way of distributing the quantum systems among the $D$ dimensions, there is one symmetric state; for two examples, see equations (\ref{threesym}) and (\ref{foursym}).
We may therefore write the optimal average success probability as 
\begin{equation}
\label{success}
p_{success}=1-{{D+N-1\choose N}D^{-N}}.
\end{equation}

This result holds also if we allow only nonentangled states in $|\Psi_N\rangle$. In this case, $|\Psi_N\rangle=| \phi _1\rangle{\otimes}| \phi _2\rangle{\otimes}...{\otimes}| \phi _N\rangle$, and we obtain 
\begin{eqnarray}
\overline{\langle P_{nonsym}\rangle}
&=&Tr\{P_{nonsym}\int|\phi_1\rangle{\otimes}|\phi_2\rangle{\otimes}...{\otimes}|\phi_N\rangle
\langle\phi_1|{\otimes}\langle\phi_2|{\otimes}...\otimes{}\langle\phi_N| d\phi_1 d\phi_2 ... d\phi_N\}\nonumber\\
&=&Tr\{P_{nonsym}\int|\phi_1\rangle\langle\phi_1|d\phi_1\int|\phi_2\rangle\langle\phi_2
|d\phi_2...\int|\phi_N\rangle\langle\phi_N|d\phi_N\}\nonumber\\
&=&Tr\{ P_{nonsym} \mathbf{1}_1\otimes \mathbf{1}_2\otimes ... \otimes\mathbf{1}_N\}={D_{nonsym}\over D_{tot}}
\end{eqnarray}
exactly as before, where $\mathbf{1}_i$ denotes the identity operator for the $i$th particle. 

There are good reasons to allow the compared pure quantum systems to be entangled with each other. Quantum comparison may be used in quantum information applications, and here one often benefits from entanglement. Also, an adversary may prepare entangled states in order to cheat, and thus we need to consider this possibility. For example, no symmetric entangled state will ever fail the universal comparison test, even if its subsystems are not in the same quantum state. 
On the other hand, entanglement with an environment has to be restricted for error-free comparison strategies to be possible. If the quantum systems that are to be compared are in mixed states, it will not always be possible to obtain unambiguous outcomes. This is because a set of identical mixed density matrices does not, in general, lie completely within the symmetric subspace, so that if the particles are found to be in the non-symmetric subspace, this does not necessarily mean that the mixed density matrices of the individual states are different. As a consequence, in the presence of noise, or entanglement with an environment, we will usually be forced to consider comparison strategies which may have errors in the results. Error-free comparison may still be possible, but only if the mixed states are restricted in some way.

\section{Detailed error-free comparison}
\label{subspaces}
When comparing the pure states of more than two quantum systems, we may also ask whether it is possible to obtain more information about the differences between their states. In this section we will show that it is possible to tell, without error, whether `all the $N$ states were  not identical', `no $N-1$ states were all identical', and so on. The  comparison strategy considered in section \ref{secuni} has only the results `all the $N$ states were not identical' and `no definite information'.
It corresponds to performing a projection onto the symmetric and non-totally symmetric subspaces.  The non-totally symmetric subspace can be further decomposed into subspaces, which also are invariant under permutations and unitary transformations. These subspaces are connected with more specified information about differences among the states.
 
To understand this, let us first illustrate this with some examples. When comparing the states of three two-dimensional quantum systems (qubits), the totally symmetric subspace is four-dimensional.  If the basis states of qubit $i$, $i=1,2,3$, are denoted by $|0\rangle_i$ and $|1\rangle_i$, a possible choice of symmetric basis states is
\begin{equation}
\label{threesym}
|0\rangle_1{\otimes}|0\rangle_2{\otimes}|0\rangle_3,\quad
|1\rangle_1{\otimes}|1\rangle_2{\otimes}|1\rangle_3,\quad
S(|0\rangle_1{\otimes}|0\rangle_2{\otimes}|1\rangle_3),\quad
S(|0\rangle_1{\otimes}|1\rangle_2{\otimes}|1\rangle_3),
\end{equation}
where $S$ denotes symmetrisation.The non-totally symmetric subspace is also four-dimensional. In this particular case, it does not decompose into further subspaces, but the case of three qubits is nevertheless useful for understanding the general situation. We may construct the non-symmetric states
\begin{equation}
|\psi_{i,j}^-\rangle \otimes|0\rangle_k, \quad|\psi_{i,j}^-\rangle \otimes|1\rangle_k,
\end{equation}
where $|\psi_{i,j}^-\rangle={1\over \sqrt{2}}(|0\rangle_i{\otimes}|1\rangle_j-|1\rangle_i{\otimes}|0\rangle_j)$ and $(i,j,k)=(1,2,3), (2,3,1)$ or (3,1,2). These six states form an overcomplete basis for the non-totally symmetric space. We have chosen this basis, rather than an orthogonal basis, because projecting onto $P_{i,j}$ and $\mathbf{1}-P_{i,j}$, with $P_{i,j}=|\psi_{i,j}^-\rangle\langle\psi_{i,j}^-|\otimes (|0\rangle_{kk}\langle 0|+|1\rangle_{kk}\langle 1|)$,
corresponds to {\it pairwise} comparison of qubits $i$ and $j$. The projector onto the non-totally symmetric subspace may be written
\begin{equation}
P_{nonsym}=\mathbf{1}-P_{sym}={2\over 3}(P_{1,2}+P_{2,3}+P_{3,1}).
\end{equation}
This can be understood to mean `in the non-symmetric subspace, at least one pair of quantum systems is different from each other', or, `at most two states are the same'.

\begin{figure}
\center{\includegraphics[width=6.5cm,height=!]{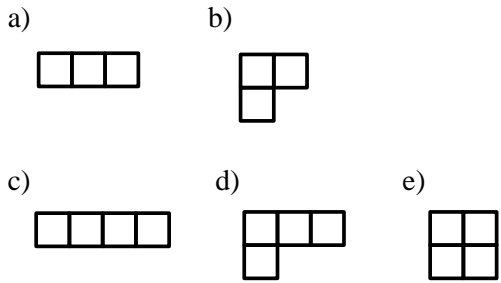}}
\caption{Young tableaux for three and four qubits. a) and b) correspond to three qubits, c), d) and e) to four qubits. a) and c) are the totally symmetric subspaces, b), d) and e) are non-symmetric.}
\label{youngfig1}
\end{figure}

When comparing four qubits, the symmetric subspace is five-dimensional with the basis states
\begin{eqnarray}
\label{foursym}
&&|0\rangle_1|0\rangle_2|0\rangle_3|0\rangle_4,\quad
|1\rangle_1|1\rangle_2|1\rangle_3|1\rangle_4,\quad\nonumber\\
&&S(|0\rangle_1|0\rangle_2|0\rangle_3|1\rangle_4),\quad
S(|0\rangle_1|0\rangle_2|1\rangle_3|1\rangle_4),\quad
S(|0\rangle_1|1\rangle_2|1\rangle_3|1\rangle_4),
\end{eqnarray}
where the tensor products have been suppressed for brevity.
It is possible to decompose the eleven-dimensional non-totally symmetric subspace into two further subspaces. These can be represented by Young tableaux \cite{youngtab} as shown in figure \ref{youngfig1}, where also the two Young tableaux for three qubits are shown. Young tableaux are a group-theoretic tool for determining the subspaces of a multi-particle system, and the dimensionalities of these subspaces. The number of boxes in a tableau is equal to the number of particles. There can be at most as many rows as there are dimensions for one quantum system, e.g. two rows for qubits. Roughly speaking, a horizontal row with $m$ boxes corresponds to $m$ particles in a symmetric state, and a vertical row with $n$ boxes correspond to $n$ particles in an antisymmetric state. 
For four qubits, the non-symmetric subspace corresponding to figure 1d) is nine-dimensional and contains states of the type
\begin{equation}
\label{1sub}
|\psi_{i,j}^-\rangle{\otimes}|0\rangle_k{\otimes}|0\rangle_l,\quad
|\psi_{i,j}^-\rangle{\otimes}|1\rangle_k{\otimes}|1\rangle_l,\quad
|\psi_{i,j}^-\rangle{\otimes}{1\over\sqrt{2}}(|0\rangle_k{\otimes}|1\rangle_l+
|1\rangle_k{\otimes}|0\rangle_l),
\end{equation}
where $(i,j,k,l)$ is any permutation of (1,2,3,4). The qubits $i$ and $j$ are in an antisymmetric state with respect to each other, and qubits $k$ and $l$ are symmetric with respect to each other.
Subspace 1e) is two-dimensional and contains the states
\begin{equation}
\label{2sub}
|\psi_{i,j}^-\rangle\otimes|\psi_{k,l}^-\rangle,
\end{equation}
where again $(i,j,k,l)$ is any permutation of (1,2,3,4). Here both qubits $i,j$ and $k,l$ are antisymmetric with respect to each other.
The two sets of states (\ref{1sub}) and (\ref{2sub}) span their respective subspaces but are overcomplete. We now see that if the four compared qubits are found in the non-symmetric subspace 1e), no selection of three qubits can have been in exactly the same state. At most two can be identical. Equivalently, all the states where three or four qubits are identical are found in the totally symmetric subspace 1c) and the first non-symmetric subspace, 1d). 
In the non-symmetric subspace 1d), no four qubits are the same, but three of them may be identically prepared. In the totally symmetric subspace, we find all sorts of combinations of states, for example, all four particles may be the in the same internal state, or the states may all be different. Therefore detection of the four qubits in the symmetric subspace gives no definite information about whether their states are same or different.

In analogy with the three-qubit case, we may write the projector onto the non-symmetric subspace as
\begin{equation}
P_{nonsym}=\mathbf{1}-P_{sym}={1\over 2}\sum_{ijkl} P_{i,j} (\mathbf{1}-P_{k,l})
+{1\over 3}\sum_{ijkl} P_{i,j}P_{k,l},
\end{equation}
where the sums are taken with $(i,j,k,l)$ equal to permutations of (1,2,3,4) with $i>j$ and $k>l$.

Let us then think about comparing the states of three three-dimensional pure quantum systems. We will be able to use what we learned from comparing three qubits, but in addition to a symmetric subspace and a non-symmetric subspace, there is now also a totally antisymmetric subspace corresponding to the Slater determinant $A(|0\rangle_1{\otimes}|1\rangle_2{\otimes}|2\rangle_3)$, where $A$ denotes antisymmetrisation. This state is antisymmetric in any pair of particles, which implies that finding the particles in this state indicates that no two of them were the same (i.e. they were all different). The non-symmetric subspace, on the other hand, corresponds to no three states being identical, which is a weaker statement.

We now know enough to understand the general case of comparing the states of $N$ $D$-dimensional quantum systems. Usually, the non-totally symmetric subspace contains only states with no definite overall symmetry. Totally antisymmetric states will exist only when the dimensionality $D$ is greater than or equal to the number of systems $N$. In a totally antisymmetric state, no two states of quantum systems are identical. In this case, there will be non-symmetric subspaces corresponding to `no $N$ states are identical', `no $N-1$ states are identical', `no $N-2$ states are identical', and so on, all the way down to `no two states are identical'. If $D<N$, there will be non-symmetric subspaces ranging from `no $N$ states are identical' only to `no $ceil[N/D]$+1 states are identical'. Here $ceil[N/D]$ denotes the smallest integer greater than or equal to $N/D$, and it is equal to the number of particles in at least one slot, when fitting $N$ particles into $D$ different slots, spreading them out as much as possible.
There may be more than one subspace corresponding to each step; the subspaces are all invariant under permutations and unitary transformations. 

\begin{figure}
\center{\includegraphics[width=7cm,height=!]{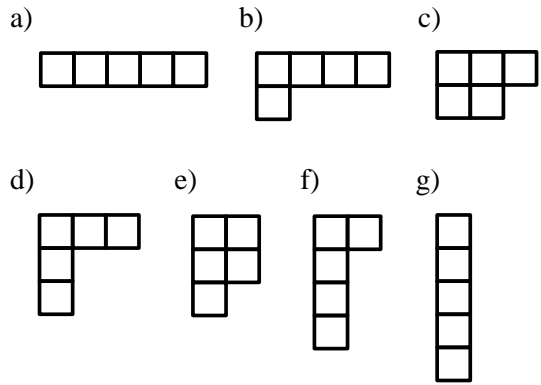}}
\caption{Young tableaux for five five-dimensional quantum systems. a) is the totally symmetric subspace, b), c), d), e) and f) are non-symmetric, and g) is the totally antisymmetric subspace.}
\label{youngfig2}
\end{figure}

The Young tableaux are the key to understanding this structure, and we give an example in figure \ref{youngfig2}. When comparing five five-dimensional quantum systems, there are seven different subspaces, corresponding to 2a), the totally symmetric subspace, 2b), a subspace where at most four particles are the same, 2c) and 2d) where at most three particles are the same, 2e) and 2f) where at most two particles are the same, and, finally, the totally antisymmetric subspace 2g) where no two particles are the same. The difference between subspaces 2c) and 2d) is that in 2c), three of the quantum systems may be identical to each other, and the remaining two in turn identical to each other, whereas in 2d), if three quantum systems are identical, the other two have to be different from each other. In 2e), there may be two identical pairs of states, whereas in 2f), only one identical pair may exist. 
This structure holds also for comparing the states of five quantum systems of higher dimension than five. The only difference is that the dimensions of the subspaces will be different, thus affecting the probabilities for the different results. When there are only four dimensions, only subspaces 2a) to 2f) may occur, when there are only three dimensions subspaces 2a) to 2e), and for five qubits, only 2a), 2b) and 2c) occur.

We have seen that it is possible to obtain more detailed information than `no, the quantum states of the particles are not all identical'. The full optimal universal error-free comparison strategy is a projection onto the different invariant subspaces of the $N$-particle system. The probabilities of the different outcomes can be calculated in analogy with equation (\ref{success}), and will be equal to $D_i/D_{tot}$, where $D_i$ is the dimension of the particular subspace considered, and $D_{tot}$ is the dimension of the total $N$-particle Hilbert space. Finding the systems in the symmetric subspace will correspond to an inconclusive result exactly as before.

\section{Minimum-error and minimum-cost comparison}
\label{prob}
In this section, we will drop the requirement that the comparison has to be error-free, and instead seek to minimise the error, or the cost of making an error. The outcome of the comparison is now either `all the systems were in the same state' (again, only the overall phase of the total state of all the systems is defined) or `the systems were not all in the same state'. There is no inconclusive result, but instead the result may sometimes be wrong. The minimum error probability will be less than if one uses an error-free comparison strategy, and then, if the result is inconclusive, guesses whether the states were identical or not. Comparing the states of the quantum systems is equivalent to distinguishing between two density matrices, $\rho_S$ for when the states are all identical, and $\rho_D$ for when the states where not all identical. The problem of distinguishing between two density matrices with minimum error was solved already by Helstrom in his pioneering work \cite{hel}. We will briefly outline the method here, applied to quantum comparison. 
Any prior information, which is available about the quantum systems and their states, is used in the construction of $\rho_S$ and $\rho_D$.
Let us denote their respective prior probabilities  by $p_S$ and $p_D$. We then form the operator
\begin{equation}
\hat{O}=p_S\rho_S-p_D\rho_D.
\end{equation}
A way to distinguish between $\rho_S$ and $\rho_D$ with least possible error is then to measure this operator, and whenever a positive eigenvalue $\lambda_i$ is obtained, take the result to be `same' ($S$). Whenever a negative eigenvalue is obtained, the result is `different' ($D$). If a zero eigenvalue occurs, one guesses randomly between `$S$' and `$D$'. In terms of measurement operators, the measurement operator $\Pi_S$ for the result `$S$' is a projection onto all eigenstates with positive eigenvalues, plus one half times the projection onto all eigenstates with zero eigenvalue. The measurement operator $\Pi_D$ for the result `$D$' is a projection onto all eigenstates with negative eigenvalues, plus one half times the projection onto all eigenstates with zero eigenvalue,
\begin{eqnarray}
\Pi_S&=&\sum_{\lambda_i>0}|\lambda_i\rangle\langle\lambda_i|+
\frac{1}{2}\sum_{\lambda_i=0}|\lambda_i\rangle\langle\lambda_i|\nonumber\\
\Pi_D&=&\sum_{\lambda_i<0}|\lambda_i\rangle\langle\lambda_i|+
\frac{1}{2}\sum_{\lambda_i=0}|\lambda_i\rangle\langle\lambda_i|.
\end{eqnarray}
The difference between $p_c$, which is the probability for the result to be correct, and the error probability $p_e$, can be expressed as
\begin{eqnarray}
p_c-p_e&=&p(S,S)+p(D,D)-p(S,D)-p(D,S)\nonumber\\
&=&\text{Tr}\left[(p_S\rho_S-p_D\rho_D)\Pi_S-(p_S\rho_S-p_D\rho_D)\Pi_D\right]\nonumber\\
&=&\sum_{i}|\lambda_i|=\text{Tr}\sqrt{(p_S\rho_S-p_D\rho_D)^2}.
\end{eqnarray}
Here $p(i,j)$ denotes the overall probability that the state is $\rho_j$ and the result obtained is $i$, and we have used the fact that the conditional probability to obtain result $i$ when the state is known to be $\rho_j$ is $\text{Tr}(\rho_j\Pi_i)$, so that $p(i,j)=p_j\text{Tr}(\rho_j\Pi_i)$.
Since $p_e+p_c=1$, the minimum error probability of the comparison is given by
\begin{equation}
p_e=\frac{1}{2}\left[1-\text{Tr}\sqrt{(p_S\rho_S-p_D\rho_D)^2}\right]=\frac{1}{2}(1-\sum_{i}|\lambda_i|).
\end{equation}
In practical applications, the cost of making errors of different kinds often varies.
When comparing quantum systems, the cost incurred when obtaining the result `different' when the states really were identical, may be very different from the cost of obtaining the result `same' when the states really were different. For example, the former case may mean throwing away a valid set of quantum systems, presented by honest participants in a quantum protocol, whereas the latter, accepting a set of non-identical states as identical, may mean that an attempt to cheat succeeds. Let us denote the cost of obtaining the result `same' when the states really were different by $C_{SD}$, and the cost of obtaining the result `different' when the states were the same by $C_{DS}$. By $C_{SS}$ and $C_{DD}$ we denote the costs for obtaining the correct results (usually these are less than or equal to zero). Instead of the error, we now want to minimise the Bayes cost
\begin{equation}
C_B=p(S,S)C_{S,S}+p(D,D)C_{D,D}+p(D,S)C_{D,S}+p(S,D)C_{S,D}.
\end{equation}
In order to do this, we have to slightly modify the minimum-error comparison procedure and instead form and diagonalise \cite{hel}
\begin{equation}
\hat{O}_B=p_S(C_{DS}-C_{SS})\rho_S-p_D(C_{SD}-C_{DD})\rho_D.
\end{equation}
Exactly as for the minimum-error comparison, positive eigenvalues of this operator correspond to the result `same' and negative eigenvalues correspond to the result `different', while zero eigenvalues correspond to a random guess. The minimum Bayes cost will be
\begin{equation}
C_B=\frac{1}{2}\left[(C_{SS}+C_{DS})p_S+(C_{DD}+C_{SD})p_D-\sum_{i}|\lambda_i|\right],
\end{equation}
where we could also use $\sum|\lambda_i|=\text{Tr}\sqrt{\hat{O}_B^2}$. Note that the choice $C_{SS}=C_{DD}=0, C_{SD}=C_{DS}=1$, corresponds to the situation where all errors are equally costly, leading to the minimum-error comparison strategy.

In this way, the optimal comparison strategy, minimising either the error or the Bayes cost, can be constructed for any comparison situation, numerically if not analytically. Notably, the obtained comparison strategy is optimal also in the presence of noise or entanglement with an environment.
The resulting strategy will be different from the error-free universal comparison strategy considered in section \ref{secuni}. What is the minimum-error comparison strategy when we have no prior information at all about the states, other than that they are pure? The distribution of possible states is flat, and this means that the prior probability for the states to be identical is zero. If minimising the error, it is therefore optimal to always directly guess that the states are different. Since this situation is perhaps not so interesting, we continue the discussion of minimum-error and minimum-cost comparison assuming that there is some kind of prior information available about the states which are compared. This prior information is used in the construction of $\rho_S$ and $\rho_D$ to obtain the optimal comparison strategy. Below we give an example to illustrate the procedure. 

\subsection{Prior knowledge about the states of the systems}
\label{prior}
Often there is prior information about the quantum systems which are compared. 
For example, it may be known that their states are identical with a certain probability $p$, and not identical with a probability $1-p$. Another example is when each state is known to be a member of a specified set of states $\{|\psi_1\rangle,|\psi_2\rangle,...,|\psi_m\rangle\}$. As an illustration of the method, we will explicitly construct the minimum-error comparison strategy for two quantum systems, when each of them are prepared in one of the so-called trine states
\begin{eqnarray}
|\psi_1\rangle&=&|+\rangle\nonumber\\
|\psi_2\rangle&=&\frac{1}{2}(-|+\rangle -\sqrt{3}|-\rangle)\nonumber\\
|\psi_3\rangle&=&\frac{1}{2}(-|+\rangle +\sqrt{3}|-\rangle),
\end{eqnarray}
with equal probability 1/3 for each of the trine states. Here $|+\rangle$ and $|-\rangle$ are orthogonal basis states.

If the states of the two quantum systems are identical, they can be either both in state $|\psi_1\rangle$, or both in state $|\psi_2\rangle$, or both in state $|\psi_3\rangle$. Each of these possibilities is equally likely, so the density matrix when the states are identical is given by
\begin{equation}
\rho_S=\frac{1}{3}(|\psi_{11}\rangle\langle\psi_{11}|+|\psi_{22}\rangle\langle\psi_{22}|+|\psi_{33}\rangle\langle\psi_{33}|),
\end{equation}
with prior probability $p_S=1/3$, where $|\psi_{ij}\rangle\equiv|\psi_i\rangle\otimes|\psi_j\rangle$. If the states are not identical, the density matrix is given by
\begin{equation}
\rho_D=\frac{1}{6}\left(|\psi_{12}\rangle\langle\psi_{12}|+|\psi_{23}\rangle\langle\psi_{23}|+|\psi_{31}\rangle\langle\psi_{31}|+|\psi_{21}\rangle\langle\psi_{21}|+|\psi_{32}\rangle\langle\psi_{32}|+|\psi_{13}\rangle\langle\psi_{13}|\right),
\end{equation}
with prior probability 2/3. An explicit calculation of $p_S\rho_S-p_D\rho_D$ and diagonalisation of this matrix  gives the eigenvalues  $\lambda_1=\lambda_2=-1/12$, $\lambda_3=-1/4$ and $\lambda_4=1/12$. We recall that negative eigenvalues correspond to the result `different', and positive eigenvalues correspond to the result `same'. The eigenvector corresponding to the only positive eigenvalue, $\lambda_4$, is $1/\sqrt{2}(|++\rangle+|--\rangle)$, and therefore the best comparison strategy is to make a projective measurement to see whether the two particles are found in this state or not. If it is, the result of the comparison is `identical', if it isn't, the result is `not identical'. The minimum error probability obtained in this way is $p_e=1/2(1-\sum|\lambda_i|)=1/4$. If we would instead use the error-free comparison strategy, which is a projection onto the symmetric versus antisymmetric subspaces, combined with a guess whenever we obtain an inconclusive outcome, it can be shown that the least error probability possible to attain is 1/3, which is larger than 1/4. (It turns out to be optimal to always guess that the states are different.)

We should note, however, that in the special case we are considering, we could combine the minimum-error and error-free comparison strategies, by making a Bell measurement as follows. When the two particles is found in the state $1/\sqrt{2}(|+-\rangle-|-+\rangle)$, we know with certainty that their states must have been different. If the result is $1/\sqrt{2}(|++\rangle+|--\rangle)$, our best guess, not necessarily correct, is that the states were identical. If $1/\sqrt{2}(|++\rangle-|--\rangle)$ or $1/\sqrt{2}(|+-\rangle+|-+\rangle)$ is obtained, the best guess is that the states were different. Strictly speaking we do not have to distinguish between all four Bell states, as we do not have to separate these last two states from each other.

Although the example given here concerns comparison of the pure states of only two quantum systems, the method is generally valid for any number of quantum systems, pure or mixed.
If we only have two outcomes of the comparison, the optimal strategy can always be constructed. These two outcomes do not have to be `all states are identical' and `not all identical', they could equally well be, for example, 'at least $M$ out of $N$ states are identical' and `at most $M-1$ states are identical'. In distinguishing between these two possibilities we follow a procedure analoguous to distinguishing between `all identical' and `not all identical'. This may be useful in a situation where we are willing to tolerate differences between the states up to a certain level.

\subsection{Detailed minimum-error and minimum-cost comparison}

As for error-free comparison, we can attempt to construct minimum-error and minimum-cost comparison strategies which will tell us more about similarities and differences among the states of the quantum systems. When comparing $N$ quantum systems, we can, for example, ask whether all states are identical, or $N-1$ states are identical, $N-2$ states identical, and so on, down to all states being different. Distinguishing between these cases is again equivalent to distinguishing between the density matrices corresponding to each of the cases. Unfortunately, the general problem of how to distinguish between more than two density matrices, with either minimum error or minimum Bayes cost, is extremely difficult to solve analytically. Even the problem of distinguishing between pure states has only been solved in a handful of special cases \cite{minerr}.
A set of conditions which the optimal solution has to satisfy is known \cite{hel}, but these conditions are of no great help in obtaining it. Nevertheless, given an explicit comparison situation, the optimal strategy could be obtained numerically by linear optimisation.

\section{Realisations with linear elements}
\label{realisations}

To experimentally realise quantum comparison may often require non-trivial manipulations, such as collective measurements on many quantum systems. An example of a collective measurement is Bell state detection, which is not possible with unit efficiency with only linear components \cite{john}.  Bell state detection corresponds to distinguishing between two-particle entangled states. Universal error-free quantum comparison corresponds to projecting onto permutation- and unitary transformation-invariant $N$-particle subspaces, which is essentially the same as distinguishing between groups of entangled $N$-particle states. Therefore realisations of quantum comparison  may well be non-trivial, essentially requiring quantum computation. 
We shall see, however, that the statistics of indistinguishable quantum systems makes it possible to implement universal quantum comparison using only linear elements and particle detectors, for example with a linear optical network and photon detectors. This has great importance for experimental realisations and practical applications of quantum comparison strategies.

Throughout this paper, we have been concerned with the situation when only one sample each of the $N$ particles is available. Experimentally it is usually easier to prepare many copies of a certain quantum system, for example, many photons in a coherent state in a laser beam, rather than a only single copy. If many, or even an unlimited number of copies of each particle are available, it is possible to estimate the state of each quantum system individually and compare their states using this information. 
In a sense, estimating the quantum systems individually resembles a `classical' comparison strategy. Here, however, we are interested in the quantum limits to comparison, and this is why we choose to consider the experimentally more difficult scenario where only one copy of each quantum system is available. We will mainly consider realisations of the error-free comparison strategy treated in section \ref{secuni}, and make some brief remarks about the realisation of minimum-error and minimum-cost comparison strategies.

\subsection{Comparing the states of two quantum systems}

A $2\times 2$ beam splitter may be used to distinguish between symmetric and antisymmetric polarisation states of two photons \cite{zeil}. The two photons will always exit in the same direction if they are in a symmetric state of polarisation, but in different directions if they are in the antisymmetric state of polarisation. In essence, this is a partial Bell state measurement. As a consequence, error-free comparison of two photonic qubits could be effected with a beam splitter, which is a linear optical element, with photon detectors at the outputs \cite{steve}. The outcome `different'  will correspond to the photons exiting in different directions (an antisymmetric state), and the inconclusive outcome will correspond to the photons exiting in the same direction. This implements the desired projection with unit probability. 

Polarised photons may here be thought of as qubits. Using different states of angular momentum, photons may also be used to represent qudits, with, in principle, arbitrary many levels per photon \cite{sonja}. We now note that the beam splitter universal comparison strategy will still be possible. This is because the non-polarising beam splitter actually affects the {\it spatial} state of the two photons. 
The beam splitter effects the transformation
\begin{eqnarray}
\label{bs2}
\hat{a}^\dagger_{0,out}&=&{1\over{\sqrt{2}}}(\hat{a}^\dagger_{0,in}+
\hat{a}^\dagger_{1,in})\nonumber\\
\hat{a}^\dagger_{1,out}&=&{1\over{\sqrt{2}}}(\hat{a}^\dagger_{0,in}-\hat{a}^\dagger_{1,in}),
\end{eqnarray}
where $\hat{a}^\dagger_{i,in}$ and $\hat{a}^\dagger_{i,out}$ are the creation operators for a particle in the spatial input and output mode $i=0,1$.  The overall state of the two photons has to be symmetric, since they are bosons. This means that a totally symmetric polarisation or angular momentum state, $|\psi_{sym}\rangle$, is connected with a totally symmetric spatial state,
$ \hat{a}_1^\dagger\hat{a}_0^\dagger |0\rangle$.  Using equation (\ref{bs2}), one sees that this spatial input state always results in the two photons exiting  together as $\hat{a}_0\hat{a}_0|0\rangle$ or  $\hat{a}_1\hat{a}_1|0\rangle$. If the photons leave in different directions, their internal state can therefore not have been symmetric, and consequently the internal states were not identical. This is why the spatial beam splitter can be used for comparing the internal states of the photons, be it polarisation or angular momentum, even if it affects the spatial part of the wave function.
The beam splitter method of comparing two photons would also work for other types of quantum systems, if a beam splitter is experimentally available. Possible physical systems include ions \cite{iontrap} or neutral atoms in traps \cite{atomtrap}. For fermions, a totally symmetric internal state is connected with a totally {\it anti}symmetric spatial state. If two fermions exit together, leaving the other exit empty, their internal states must clearly have been different --- no two fermions may occupy the same quantum state. 

\subsection{Comparing the states of more than two quantum systems}
For two quantum systems, a beam splitter is enough to implement universal error-free quantum comparison with unit efficiency. There is, however,  no immediate reason why linear elements would be sufficient for error-free comparison of the states of more than two quantum systems. The fact that we are asking for unambiguous results complicates the situation. A careful investigation, however, shows that linear elements are enough, at least if one is satisfied with a difference detection probability which may be less than ideal, but still nonzero. One now has to use a $N\times N$ beam splitter, which can be implemented with a network of $2\times 2$ beam splitters \cite{igornetwork}. The $N\times N$ beam splitter transform affects only the spatial part of the quantum state and may be written as
\begin{equation}
\label{bstrafo}
\hat{a}^\dagger_{j,out}={1\over{\sqrt{N}}}\sum_{k=0}^{N-1}
\exp(i{{2\pi}\over N}kj)\hat{a}^\dagger_{k,in}
\end{equation}
where $\hat{a}^\dagger_{j,out}$ and $\hat{a}^\dagger_{k,in}$ are the creation operators for a particle in spatial output mode $j$ and input mode $k$, respectively.
The compared particles are fed into the beam splitter, each into one input port. If their internal states are all identical, the input state is given by
\begin{equation}
\label{bsinstate}
|\Psi_{in}\rangle = \hat{a}^\dagger_{0,in} \hat{a}^\dagger_{1,in}...\hat{a}^\dagger_{N-1,in}|0\rangle_{spatial}\otimes 
|\psi\rangle_{internal}^N
\end{equation}
where $|\psi\rangle_{internal}$ is the internal quantum state of each of the $N$ particles we wish to compare, and the spatial part of the state is symmetric (bosons) or antisymmetric (fermions).  The beam splitter does not affect the internal states of the particles. For fermions in identical internal states, we thus see that both the total spatial input and output states have to be antisymmetric, with no more than one fermion in each beam splitter output port. This implies that as soon as two or more of the fermions exit in the same path, or, equivalently, that there is no `click' in some other exit, all the fermions cannot have had identical internal states. A brief consideration shows that, if only $M$ out of the $N$ detectors fire, then at most $M$ of the fermions can have been identical. 

For more than two bosons, such as photons, the click patterns indicating differences among the particles will be slightly different. 
For three bosons, a  calculation using equations (\ref{bstrafo}) and (\ref{bsinstate}) shows that, if their internal states are identical, the output state will be
\begin{equation}
\label{threebos}
{1\over {3\sqrt{3}}}(\hat{a}^{\dagger 3}_{1,out}+\hat{a}^{\dagger 3}_{2,out}+
\hat{a}^{\dagger 3}_{3,out} - 3\hat{a}^\dagger_{1,out}\hat{a}^\dagger_{2,out}\hat{a}^\dagger_{3,out})|0\rangle
\otimes|\psi\rangle ^3_{internal}
\end{equation}
(remember that $\hat{a}^{\dagger n}|0\rangle = \sqrt{n!}|n\rangle$).
All the bosons will either leave from different outputs, or all from the same. Two bosons leaving from the same output, and the third from a different output, is not possible. Therefore, when it occurs, a `two detectors fire, the third does not' click pattern indicates a difference among the internal states of the bosons. For four identical bosons, the output state will be (suppressing the `out' indices)
\begin{eqnarray}
\label{fourbos}
&&{1\over {16}}\left(\hat{a}^{\dagger 4}_{1}+\hat{a}^{\dagger 4}_{2}+\hat{a}^{\dagger 4}_{3}+ \hat{a}^{\dagger 4}_{4} 
- 2\hat{a}^{\dagger 2}_{1}\hat{a}^{\dagger 2}_{3} +
2\hat{a}^{\dagger 2}_{2}\hat{a}^{\dagger 2}_{4} \right.\nonumber\\
&&\left.-4\hat{a}^{\dagger 2}_{1}\hat{a}^\dagger_{2}\hat{a}^\dagger_{4}
+4\hat{a}^{\dagger 2}_{2}\hat{a}^\dagger_{1}\hat{a}^\dagger_{3}
-4\hat{a}^{\dagger 2}_{3}\hat{a}^\dagger_{2}\hat{a}^\dagger_{4}
+4\hat{a}^{\dagger 2}_{4}\hat{a}^\dagger_{1}\hat{a}^\dagger_{3}\right)|0\rangle
\otimes|\psi\rangle ^4_{internal}.
\end{eqnarray}
From this expression we conclude, for example, that one of the many click patterns indicating a difference is when all four bosons exit through different outputs, since this is not possible when all internal states are identical. To find out exactly which click patterns indicate differences among the internal states of $N$ bosons is a straightforward but somewhat labour-intensive task. One of these click patterns will always be $N-1$  bosons exiting from the same output port, and the remaining one from a different output port.  
A feasible option for comparing more than two photons, or other bosons,  is also to use pairwise comparison. When comparing three photons, we may compare only two of them, when comparing four photons, compare them as two pairs etc.. If at least one pair of photons are different from each other, then obviously all the photons cannot have been identical. This method will, however, be less efficient.

If we only are able to count photons at the individual exits, the efficiency of the beam splitter realisation will not, in general, be as good as for a perfect projection onto different subspaces as described in sections \ref{secuni} and \ref{subspaces}. To reach the optimal efficiency, we would have to be able to project onto states like (\ref{threebos}) and (\ref{fourbos}) and the state spaces orthogonal to these. If we would be able to measure the relative phases of the detected particles, this would improve the efficiency, but would still not be enough to reach the optimum. Whether the beam splitter realisation is optimal, given that we only can use linear elements and particle detectors, is an open question. It is, however, clear that the optimal realisation must use a balanced multiport, since the optimal realisation should again be invariant under permutations and unitary transformations. 

If we are not able to count the photons in each output detector, only tell whether a given detector is firing or not, this will further degrade the efficiency. We will then be unable to distinguish between events of the type `three photons in detector A and one photon in detector B' (indicating, say, a difference) and `two photons each in detector A and B' (say, indicating an inconclusive outcome). We will have to count both these events as `detectors A and B fired', giving an inconclusive result.
Noise, dark counts and missed detections will mean that the experimentally realised universal quantum comparison is not completely error-free anymore. The compared quantum states will also be destroyed unless we can detect in which path the individual particles exit without disturbing them. 
But the important  and perhaps somewhat surprising conclusion remains, that it is possible to  compare the quantum states of arbitrarily many quantum systems, such as photons, ions or neutral atoms, using only a linear network of beam splitters and particle detectors. 

\subsection{Realising minimum-error and minimum-cost comparison}

In section \ref{prob}, we saw that the minimum-error and the minimum Bayes cost comparison strategies always are projective measurements in some basis. Thus these comparison strategies can always be realised in principle, but as for the universal  error-free comparison, it may require a measurement in a highly entangled basis, making the realisation difficult in practice.
We gave an example of a comparison strategy for determining, with minimum error, whether the states of two quantum systems were identical or not. The states of the each of the systems were known to be in one of three given quantum states. In the particular case we considered, the quantum comparison can be realised with a beam splitter in a way similar to the universal error-free comparison. We need to distinguish the state $1/\sqrt{2}(|++\rangle+|--\rangle)$ from the other Bell states, as this state corresponds to the outcome `same' and the other Bell states (or linear combinations of these) correspond to the outcome `different'. One way to do this is to apply a rotation of $\pi/2$ to one of the quantum systems, so that $|+\rangle\leftrightarrow |-\rangle$, and $1/\sqrt{2}(|++\rangle+|--\rangle)$ transforms into $1/\sqrt{2}(|+-\rangle+|-+\rangle)$. If two quantum systems in this state are incident on a beam splitter, both systems will exit together since the state is symmetric. But this case can readily be distinguished from the other symmetric states, $1/\sqrt{2}(|++\rangle\pm|--\rangle)$, since it is the only one where the quantum systems have different states when each of them is measured in the basis $\{|+\rangle,|-\rangle\}$. To do this, direct each of the beam splitter outputs onto a polarising beam splitter, separating $|+\rangle$ and $|-\rangle$, and look for clicks at both outputs of one of the polarising beam splitters. Note that since the comparison situation is not invariant under unitary transforms, we need to know in which basis to perform the measurement. Essentially, this method has already been used to individually distinguish two Bell states from each other and from the other two Bell states \cite{zeil}.

The given example is only an illustration and does by no means have any general significance.
Optimal minimum-error and minimum-cost comparison strategies are highly dependent on the information which is available about the states and the quantum systems, and therefore also their realisations have to be individually tailored.

\section{Discussion and conclusions}

In this paper we have considered comparison of the states of quantum systems, when only one copy of each quantum system is available. We have treated error-free comparison strategies, which give definite answers, at the expense of sometimes obtaining an inconclusive result, and  minimum-error and minimum-cost comparison strategies, which sometimes may give the wrong result. Error-free strategies are useful whenever it is important that the answers obtained are guaranteed to be correct. Here a restriction to pure states is made, because for mixed quantum states, unambiguous answers will in general not be possible. In this case, one can instead opt for  a minimum-error or minimum-cost comparison strategy. Error-free comparison of mixed states may be possible, but only if  there is prior information about the mixed states.

The optimal universal error-free quantum comparison strategy, treated in sections \ref{secuni} and \ref{subspaces}, was found to be a projection onto different subspaces, invariant under permutations and unitary transformations of the quantum systems. Finding the state of the systems outside the totally symmetric subspace corresponds to the states being not all identical to each other. Finding the systems in the symmetric subspace corresponds to an inconclusive outcome, since in this case we cannot be sure if their states were all identical or not. By further checking in which subspace of the non-symmetric space the state of the systems is found in, one can obtain more detailed information about the differences and similarities among the quantum systems. For example, finding the particles in the totally antisymmetric subspace means that no two quantum systems were in the same state. This outcome is possible only if the  dimensionality $D$ of the Hilbert space for each quantum particle is greater than the number of particles $N$.

Sometimes we may have prior information about the quantum systems we want to compare. In this case, it is often possible to improve the success probability of the comparison by using this information in designing the comparison strategy. In particular, it is sometimes possible to obtain also an unambiguous answer that the states of all the systems are identical. If the states of the compared quantum systems are known to belong to a given set of states, this will be possible if and only if there is at least one state in this set which is linearly independent of the other states in the set \cite{tonyprivate}. This requires an error-free comparison strategy different from the universal strategy, depending on the individual situation.  

In section \ref{prob}, we explained how to construct the optimal comparison strategies minimising either the error probability, or the Bayes cost, which is the average cost of making an error. The resulting comparison strategy is optimal also when the states of the systems are mixed, when there is noise, or entanglement with an environment. The presented method could be used to obtain the optimal comparison strategy for applications such as quantum fingerprinting \cite{buhr} and quantum digital signatures \cite{chuang}. In these cases, there is prior information about the compared quantum states; essentially we have to compare codeword states to each other. The c-SWAP comparison strategy used in \cite{buhr, chuang} is actually one special realisation of the universal error-free comparison strategy described in section \ref{secuni}, with the difference that the inconclusive result is taken to mean that the states are identical. This way one obtains a comparison strategy with one-sided error. As we have shown, however, one can usually construct a minimum-error comparison strategy with less error than this. It is possible to use the universal comparison strategy, and this strategy may even be good, depending on the situation, but not necessarily optimal. An optimal strategy would of course improve the success rate of the protocols, but is also of importance to consider when designing strategies for cheating. 

We found, in section \ref{realisations},  that  error-free (and sometimes also minimum-error) quantum comparison can be realised using only beam splitters and particle detectors also for more than two quantum systems, making it possible to implement using linear optics. Quantum comparison could be realised this way also for any other physical systems, as soon as beam splitters and particle detectors are available. This is of great importance for practical applications.  
The c-SWAP realisation, in contrast, requires rather non-trivial quantum computation. A recent suggestion for implementing the c-SWAP strategy on two polarised photons with linear optics, besides being more complicated than using a beam splitter, succeeds only with probability 1/8 \cite{grudka}. In contrast, the beam splitter realisation  succeeds in implementing the desired projection with unit probability for two photons. The drawback is of course that the photons are absorbed by detectors, whereas the c-SWAP realisation retains the quantum states in the case when they are identical.

 When comparing more than two quantum systems, the efficiency of the beam splitter realisation is somewhat lower than ideal. It is an open question whether it is possible to reach a higher efficiency, given that only linear elements and particle detectors are available for the realisation. Relatedly, the optimal efficiency of a Bell state measurement, realised with only linear elements, is 1/2, and this is indeed reached with a beam splitter \cite{john}. To reach higher efficiencies in  a Bell state measurement requires the use of nonlinear effects, or additional entanglement in auxiliary degrees of freedom. 
We should also note that recent results on measuring the overlap of two states \cite{olomouc}  turn out to be closely related to state comparison for two states, in particular through the beam splitter realisation.

There exists a relation between quantum comparison and quantum state discrimination~\cite{hel, tonyrev}, which is currently a widely studied problem. In a state discrimination situation, one is given a quantum system, whose state belongs to some set of given quantum states, with associated prior probabilities. The task is then to determine which one of these states the system is in. The procedure may be optimised with respect to different criteria, for example minimising the error in the result. Finding optimal state discrimination strategies for a given set of possible states is usually a highly non-trivial problem.
Comparing the states of a set of given qantum systems can be rephrased as a state discrimination problem: We  have to discriminate the case when the states of the quantum systems are the same from the case when they are different. In fact, this argument is explicitly used in section \ref{prob} to obtain the minimum-error and mimimun-cost comparison strategies, and is essentially underlying the discussion also of the error-free comparison strategies in sections \ref{secuni} and \ref{subspaces}.

Also related to quantum comparison is quantum template matching~\cite{sas1, sas2}. Here the task is to determine which one of a set of given template states is closest to an unknown input state.

Quantum comparison may be used in many applications in quantum information, for example in error correction schemes, where it could be used to detect whether the results of many runs of the same quantum calculation agree or not. It could also be used to detect differences between a quantum state and one or many reference states.
To conclude, we have seen that it is possible, given only one copy each of $N$ quantum systems, to compare the states of these quantum systems.
Quantum comparison gives collective information about the states of the compared quantum systems, without obtaining individual information about their states.

\section*{Acknowledgments}
We want to thank Prof. Stephen M. Barnett for valuable discussions.
Financial support for IJ by the Ministry of Education of the Czech republic (MSMT 210000018), GACR 202/01/0318, for EA  by the EU Marie Curie programme, project number HPMF-CT-2000-00933, and for AC by the Engineering and Physics Research Council EPSRC, and the University of Hertfordshire, is gratefully acknowledged.

\end{document}